\documentclass[reprint,superscriptaddress,showpacs,preprintnumbers,amsmath,amssymb,aps,prl,]{revtex4-1}

\usepackage{bm}
\usepackage{graphicx}
\usepackage{epstopdf}
\usepackage{array} 
\usepackage{listings}
\usepackage{dcolumn}
\usepackage{booktabs}
\usepackage[colorlinks,citecolor=blue]{hyperref}

\usepackage{multirow,bigdelim}
\usepackage{float}
\usepackage{xcolor}
\usepackage{soul}
\usepackage[colorlinks,citecolor=blue]{hyperref}

\usepackage{diagbox}
\newcolumntype{C}[1]{>{\centering\let\newline\\\arraybackslash\hspace{0pt}}m{#1}}
\AtBeginDocument{
	\heavyrulewidth=.08em
	\lightrulewidth=.05em
	\cmidrulewidth=.03em
	\belowrulesep=.65ex
	\belowbottomsep=0pt
	\aboverulesep=.4ex
	\abovetopsep=0pt
	\cmidrulesep=\doublerulesep
	\cmidrulekern=.5em
	\defaultaddspace=.5em
}

\newcommand{\hfb}{UNEDF1$_{\mathrm{HFB}}$}
\newcommand{\Og}{$^{294}_{118}$Og$_{176}$}
\newcommand{\Pb}{$^{208}_{\hphantom{2}82}$Pb$_{126}$}
\newcommand{\Kr}{$^{86}_{36}$Kr$_{50}$}
\newcommand{\Xe}{$^{126}_{\hphantom{1}54}$Xe$_{72}$}

\newcommand{\MGCMp}{$\mathcal{M}^{\rm GCM}$}
\newcommand{\MATDHFp}{$\mathcal{M}^{\rm AP}$}
\newcommand{\MATDHF}{$\mathcal{M}^{\rm A}$}

\newcommand{\gras}[1]{\boldsymbol{#1}}

\begin{document}


\title{Cluster radioactivity of \Og{}}

\author{Zachary Matheson}
\affiliation{Department of Physics and Astronomy and FRIB Laboratory, Michigan State University, East Lansing, Michigan 48824, USA}

\author{Samuel Giuliani}
\affiliation{Department of Physics and Astronomy and FRIB Laboratory, Michigan State University, East Lansing, Michigan 48824, USA}

\author{Witold Nazarewicz}
\affiliation{Department of Physics and Astronomy and FRIB Laboratory, Michigan State University, East Lansing, Michigan 48824, USA}

\author{Jhilam Sadhukhan}
\affiliation{Variable Energy Cyclotron Centre, Kolkata 700064}
\affiliation{Homi Bhabha National Institute,~Mumbai 400094, India}

\author{Nicolas Schunck}
\affiliation{Nuclear and Chemical Science Division, Lawrence Livermore National Laboratory, Livermore, California 94551, USA}

\date{\today}

\begin{abstract}

According to theory, cluster radioactivity becomes an important decay mode in
superheavy nuclei. In this work, we predict that the strongly-asymmetric fission,
or cluster emission, is in fact the dominant fission channel for \Og{}, which is
currently the heaviest synthetic isotope known. Our theoretical approach
incorporates important features of fission dynamics, including quantum tunneling
and stochastic dynamics up to scission. We show that, despite appreciable
differences in static fission properties such as fission barriers and
spontaneous fission lifetimes, the prediction of cluster radioactivity in \Og{}
is robust with respect to the details of calculations, including the choice of
energy density functional, collective inertia, and the strength of the
dissipation term.  

\end{abstract}

\maketitle


{\it Introduction} -- The region of superheavy nuclei ($Z\geq104$) is one of the
frontiers of modern nuclear physics. The heavy-ion fusion  experiments have been
able to push the boundaries of the nuclear chart all the way to \Og{}
\cite{Oganessian2006,Oganessian2012,Brewer2018}, and new efforts are underway to
increase production rates of superheavy systems
\cite{Dmitriev2016,Oganessian2016,Hoffman2016,Roberto2018}. Due to the large
number of nucleons, these nuclei push the limits of nuclear structure models and
are expected to answer key questions pertaining to nuclear and atomic physics,
astrophysics, and chemistry
\cite{Duellmann2018,Jerabek2018,Nazarewicz2018,Giuliani2019}. For instance,
since the liquid drop model predicts vanishing fission barriers for superheavy
elements due to the strong Coulomb repulsion, shell effects become absolutely
essential and spontaneous fission ends up governing the lifetimes of many of
these new systems \cite{Hessberger2017,Baran2015}.  The fission of superheavy
elements may also play an important role in the astrophysical \textit{r}
process, by placing an endpoint on neutron capture and through the fission
cycling \cite{Giuliani2018}.

In the superheavy research enterprise, theory plays a critical role by guiding
experiments, interpreting their results, and making predictions in the regions
that cannot be reached  experimentally \cite{Nazarewicz2018,Giuliani2019}
because of huge proton and neutron numbers involved. In these extreme regions,
it is important to use  models of spontaneous (or low-energy) fission rooted
firmly in many-body quantum mechanics. To this end, microscopic models based on
density functional theory (DFT) where collective dynamics is decoupled from
underlying intrinsic excitations offer a path which is computationally
tractable, while still maintaining a direct link to the underlying quantum
many-body problem \cite{Schunck2016}. These models can be used to predict
observables such as half-lives \cite{Erler2012, Staszczak2013, Giuliani2013,
Giuliani2014, Schunck2016, Lemaitre2018, Rodriguez2018} and primary fragment
yields \cite{Sadhukhan2016,Warda2018,Regnier2016,Regnier2018} within the same
theoretical framework.

In this work, we discuss an exotic form of cold highly-asymmetric fission, known
in the literature as cluster radioactivity or cluster emission
\cite{Sandulescu1980,Poenaru1986,Royer1998}, which we predict is the dominant
form of fission in the superheavy isotope \Og{}. Cluster emission, an
intermediate process between $\alpha$ decay and conventional fission with
fragments of more comparable masses, occurs when a parent nucleus decays into a
large fragment near doubly-magic \Pb{} and a lighter cluster. It has been
observed experimentally in actinides, starting with the
$^{223}_{\hphantom{2}88}$Ra$\rightarrow$$^{209}_{\hphantom{2}82}$Pb +
$^{14}_{\hphantom{1}6}$C decay \cite{Rose1983}. It is always a rare event with a
small branching ratio \cite{Poenaru2010}.  From the theoretical point of view,
half-life calculations based on semiempirical models predict cluster
radioactivity to be the dominant decay channel of several superheavy
nuclei~\cite{Poenaru2011, Poenaru2012, Poenaru2013, Poenaru2015, Poenaru2018,
Santhosh2018, Zhang2018}. Similar predictions have been obtained by more
microscopic calculations using nuclear DFT framework~\cite{Warda2011,Warda2018}.
However, so far no determination of fission yields has been made that explicitly
demonstrates the emergence of cluster radioactivity in superheavy elements,
either theoretically or experimentally. In this paper, which extends the
discussion of recent Ref.~\cite{Warda2018} to fission yields, we calculate
spontaneous fission yields of \Og{} and explicitly predict for the first time
that cluster emission is dramatically enhanced to the point that it becomes the
primary spontaneous fission channel. 

Our paper is organized as follows.  First, we briefly sketch the microscopic
DFT+Langevin framework used to calculate fission fragment distributions. We then
compute the spontaneous fission characteristics of \Og{}. Finally, we study the
robustness of our fission yield predictions with respect to different energy
density functionals (EDFs), collective spaces, collective inertias, and
dissipation.


{\it Model} -- We calculate fission fragment distributions following the
approach described in Ref.\cite{Sadhukhan2016}, which may be divided into two
stages. In the first stage, we use the semiclassical WKB approximation to model
spontaneous fission as quantum tunneling through a multidimensional potential
energy surface (PES) characterized by $N$ collective coordinates
$\mathbf{q}\equiv(q_1, \ldots, q_N)$. In our implementation of the WKB
approximation, the most-probable tunneling path $\left. L(s) \right|_{s_{\rm
in}}^{s_{\rm out}}$ in the collective space is found via minimization of the
collective action
\begin{equation}\label{eq:action} 
  S(L) = \frac{1}{\hbar}\int_{s_{\rm in}}^{s_{\rm out}} \sqrt{2\mathcal{M}(s)\left(V(s)-E_0\right)}ds,
\end{equation} 
where $s$ is the curvilinear coordinate along the path $L$, $\mathcal{M}(s)$ is
the collective inertia \cite{Sadhukhan2013} and $V(s)$ is the potential energy
along $L(s)$. $E_0$ stands for the collective ground-state energy. The dynamic
programming method \cite{Baran1981} is employed to determine the path $L(s)$.
The calculation is repeated for different outer turning points, and each of
these points is then assigned an exit  probability $P(s_{\rm
out})=[1+\exp{(2S)}]^{-1}$ \cite{Baran1978}. 

In the second stage, fission trajectories begin from the outer turning
line and then evolve along the PES according to the Langevin equations:
\begin{align} 
  \frac{dp_i}{dt} = & 
  -\frac{p_j p_k}{2} \frac{\partial}{\partial q_i}\left(\mathcal{M}^{-1}\right)_{jk} 
  - \frac{\partial V}{\partial q_i} 
  \nonumber \\ 
  & -\eta_{ij}\left(\mathcal{M}^{-1}\right)_{jk} p_k + g_{ij}\Gamma_j(t) \,, \\ 
  \frac{dq_i}{dt} = &
  \left(\mathcal{M}^{-1}\right)_{ij} p_j \,, \nonumber 
\end{align} 
where $p_i$ is the collective momentum conjugate to $q_i$. The dissipation
tensor $\eta_{ij}$ is related to the random force strength $g_{ij}$ via the
fluctuation-dissipation theorem, and $\Gamma_j(t)$ is a Gaussian-distributed,
time-dependent stochastic variable. All trajectories ending at a particular
scission configuration are weighted with the appropriate $P(s_{\rm out})$.
These scission configurations were defined based on the expectation value of the
neck operator $Q_N = e^{-{(z -z_n)^2}/{a_n^2}}$, setting $a_n = 1$ fm and with
$z_n$, taken to be the point in the nucleus with the lowest density, describing
the position of the neck. We defined our scission line differently for each functional that was used, with the value $Q_N = 7$
for \hfb{} and SkM* and $Q_N = 9$ for D1S. Particle number fluctuations in the
neck at or near the scission line were accounted for by convoluting our Langevin
yields with a Gaussian function of width $\sigma_A=6$ for $A$ and $\sigma_Z=4$
for $Z$.

The key ingredients in these calculations, $V$ and $\mathcal{M}$, are calculated
self-consistently by solving the Hartree-Fock-Bogoliubov equations employing
Skyrme and Gogny EDFs. To evaluate the robustness of our results with respect to
different inputs, we perform calculations using three distinct EDFs: \hfb
\cite{Schunck2015}, a Skyrme functional which was optimized to data for
spherical and deformed nuclei, including fission isomers; SkM*
\cite{Bartel1982}, another Skyrme functional designed for fission barriers and
surface energy; and D1S~\cite{Berger1989}, a parametrization of the finite-range
Gogny interaction fitted on fission barriers of actinides.

In self-consistent fission models, the lowest multipole moments characterizing
nuclear shape deformations  are usually selected as collective coordinates. The
remaining shape degrees of freedom are, in principle, decided through the energy
minimization. In the present work, axial quadrupole moment $Q_{20}$, triaxial
quadrupole moment $Q_{22}$, and axial octupole moment $Q_{30}$ are considered as
collective coordinates since the fission dynamics associated with fragment-yield
distributions is mostly confined  within this deformation space. Additionally,
pairing correlations have a strong impact on the spontaneous fission half-lives
calculated via action minimization \cite{Sadhukhan2014,Giuliani2014,Zhao2016}.
It is taken into account through the coordinate $\lambda_2$ representing dynamic
pairing fluctuations \cite{Sadhukhan2014}. To obtain $S(L)$,  a dimensionless
collective space is introduced as in Ref.~\cite{Sadhukhan2014}. 

To balance computational speed with complexity, we used a different collective
space for each functional. The most detailed calculation was carried out using
\hfb{} in a four dimensional space $(Q_{20},Q_{22},Q_{30},\lambda_2)$.
Calculations were performed using the symmetry-unrestricted DFT solver HFODD
\cite{Schunck2017}.  To assure good convergence,  we used the 1500 lowest single
particle levels corresponding to  30 stretched harmonic oscillator shells.

The analysis of the four-dimensional (4D) PES  showed that $Q_{30}$ remains
negligible through the first saddle point (up to at least $Q_{20}=100$\,b), and
that $Q_{22}$ and $\lambda_2$ are unimportant beyond the outer turning-point
hyper-surface. This allowed us to  simplify calculations with  other
functionals. The SkM* calculations were performed in a piecewise space
($(Q_{20},Q_{22},\lambda_2)$ up to the fission isomer,
$(Q_{20},Q_{30},\lambda_2)$ from fission isomer to outer turning points, and
$(Q_{20},Q_{30})$ beyond the outer turning-point line) with the same pairing
properties as given in \cite{Mcdonnell2014,Sadhukhan2013,Sadhukhan2014}, and the
same HFODD basis  as in \hfb{} calculations. The PES connection at the fission
isomer assumed $Q_{22}=Q_{30}=0$, with $Q_{20}$ and $\lambda_2$ continued
smoothly, as in \cite{Sadhukhan2016}. In the case of Gogny D1S calculations, a
two-dimensional collective space described by coordinates $(Q_{20},Q_{30})$ was
used within the DFT solver HFBaxial \cite{Robledo2002}, where the stretched
harmonic oscillator basis corresponding to 17 harmonic oscillator shells was
optimized for each $(Q_{20},Q_{30})$ value. Each PES was interpolated from a
discrete, uniform grid, the spacing of which is shown in the upper half of
Table~\ref{tab:PES}.

\begin{table}[!htb]
	\caption{\label{tab:PES}
	Top: Mesh sizes used for the discretization of the potential energy surface.
	Bottom: Step sizes used in the calculation of the non-perturbative inertia \MATDHF.}
\begin{ruledtabular}
	\begin{tabular}{lcccccc}
		&                     \multicolumn{4}{c}{WKB region}                               &  \multicolumn{2}{c}{Langevin region} \\
		\cmidrule(r){2-5} \cmidrule{6-7} 
		&  $\Delta Q_{20}$  &  $\Delta Q_{22}$  &  $\Delta Q_{30}$  &  $\Delta \lambda_2$  &  $\Delta Q_{20}$ &  $\Delta Q_{30}$  \\
		&       (b)         &         (b)       &    (b$^{3/2}$)    &    (MeV)             &          (b)     &     (b$^{3/2}$)   \\
		
		\midrule
		\hfb{} &        6          &          3        &        3          &     0.05             &           4      &        2           \\
		SkM*   &        4          &          -        &        2          &     0.05             &           4      &        2           \\
		D1S    &        4          &          -        &       2.5         &        -             &           4      &       2.5          \\
		\midrule
		\hfb{} &        6          &          3        &        3          &     0.05             &         0.001    &      0.001         \\
		SkM*   &        4          &          -        &        2          &     0.05             &         0.001    &      0.001         \\
		D1S    &       0.1         &          -        &       0.1         &        -             &          0.1     &       0.1          \\
	\end{tabular}
	\end{ruledtabular}
\end{table}

Several approximations are commonly used to compute the collective inertia,
which describes the tendency of the nucleus to resist configuration changes.
The most accurate prescription available to date is obtained from
non-perturbative cranking approximation to the adiabatic time-dependent
Hartree-Fock-Bogoliubov (ATDHFB) inertia ({\MATDHF}) \cite{Baran2011}.  On the
other hand, perturbative expressions prioritize computational simplicity by
sacrificing details of the level crossing dynamics, which results in a
smoothed-out collective inertia. We also performed calculations using both the
perturbative cranking ATDHFB inertia ({\MATDHFp}) \cite{Baran2011} and the
perturbative GCM inertia ({\MGCMp}) \cite{Schunck2016}, which has the same
structure as {\MATDHFp} but with an absolute magnitude quenched by a factor
1.5~\cite{Giuliani2018b}. In \hfb{} and SkM* the derivatives were computed by
means of the Lagrange three-point formula for unequally spaced points (see
Ref.~\cite{Baran2011}), while first order finite difference formulas were used
for D1S (see Ref.~\cite{Giuliani2018b}). A summary of the step sizes used in the
calculation of \MATDHF{} for each collective variable and region of the PES is
given in the lower half of Table~\ref{tab:PES}.

In this work, we set the dissipation strength $\eta_{ij}$ as an adjustable
parameter rather than using some of the common prescriptions
\cite{usang2017,ishizuka2017}, which have not yet been adapted to DFT inputs. We
examined the sensitivity of our calculations to $\eta_{ij}$  by varying it
around the baseline value used in \cite{Sadhukhan2016}: $\gras{\eta}_0 \equiv
\left(\eta_{22},\eta_{23},\eta_{33}\right)\equiv(50\hbar, 5\hbar, 40\hbar)$.

\begin{figure} [htb]
\includegraphics[width=1.0\linewidth]{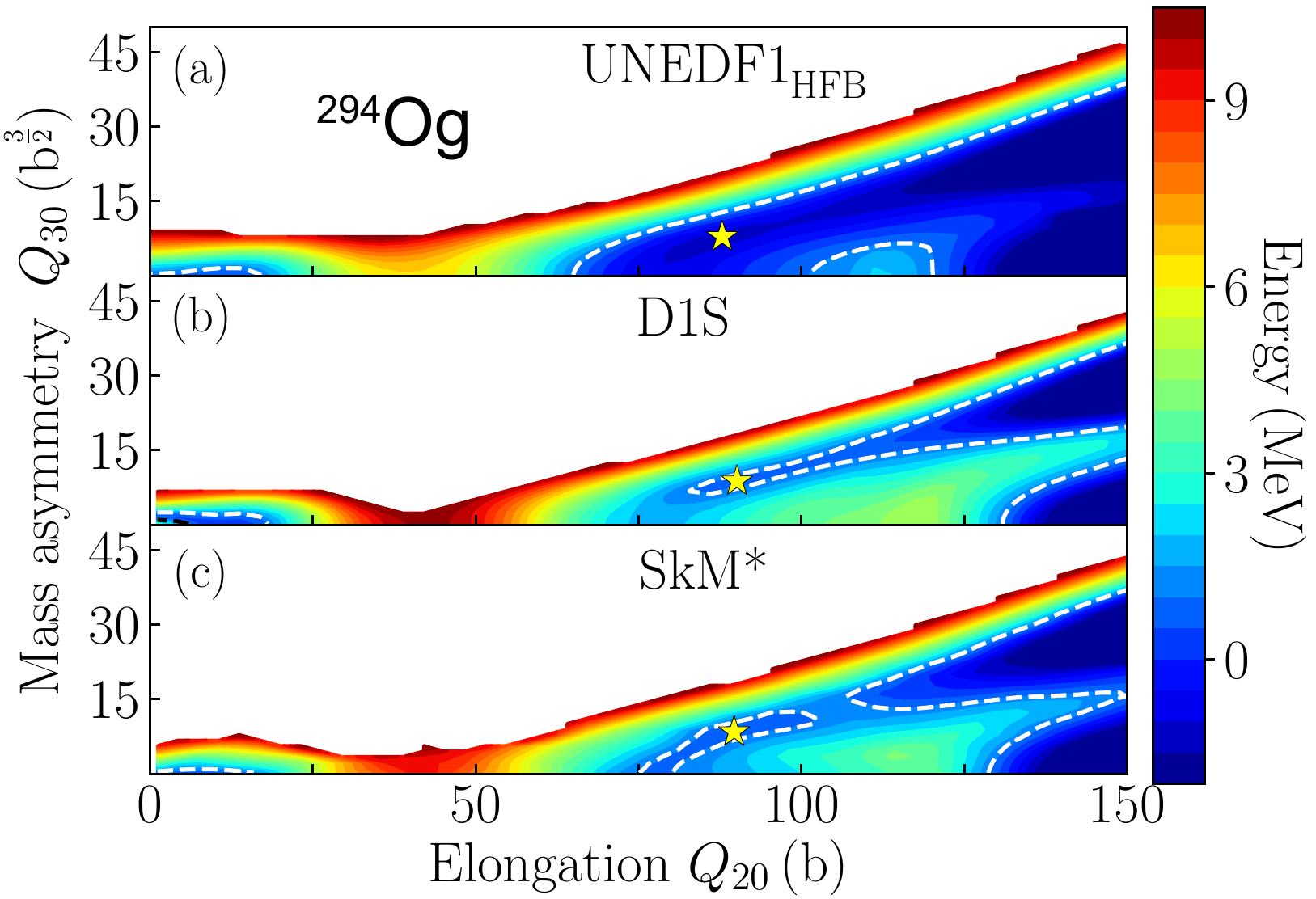} 
        \caption[\Og{} PES from \hfb{}, SkM* and D1S]{
	Comparison of the PESs for \Og{} in the $(Q_{20},Q_{30})$ collective
	plane obtained in  \hfb{} (a), D1S (b), and SkM* (c) EDFs. The
	ground-state energy $E_{gs}$ is  normalized to zero. The dotted line in
	each figure corresponds to $E_0-E_{gs}=1$\,MeV, which was used to
	determine the inner and outer turning points. The local energy minima at
	large deformations are marked by stars.}
	\label{fig:pes} 
\end{figure}

{\it Results} -- We first compare in Fig.~\ref{fig:pes} the  two-dimensional
PESs in the $(Q_{20},Q_{30})$ plane for the functionals \hfb{}, D1S, and SkM*.
There we notice that the overall topology of the PES is roughly similar in all
models, with a symmetric saddle point occurring around $Q_{20} \approx 40$\,b, a
second barrier beginning around $Q_{20}\approx100-120$\,b along the symmetric
fission path, the presence of local minima at large deformations, a deep valley
that leads to an highly-asymmetric split, and the secondary less-asymmetric
fission valley that emerges at large elongations.

But there are  differences as well, such as the height of the first saddle
point, the depth of the highly-asymmetric fission valley, and the height of the
ridge  separating the two fission valleys. As a result,  the outer turning
points are pushed to larger elongations in D1S and SkM* as compared to \hfb{}.
These differences in the PES topology strongly affect the predicted spontaneous
fission half-lives $\tau_\mathrm{SF}$, which in the case of \hfb{}, SkM* and D1S
are $9.1\times10^{-9}\,$s, $4.0\times10^{-5}\,$s and $3.2\times10^{-2}\,$s,
respectively (see also \cite{Staszczak2013,Baran2015} for a detailed discussion
of half-lives). These large variations of $\tau_\mathrm{SF}$ reflect the
well-known exponential sensitivity of spontaneous fission half-lives to changes
in the quantities entering the collective action \eqref{eq:action}. The
$\tau_\mathrm{SF}$ predictions of \hfb{} and, to a lesser degree,  SkM*  are
incompatible with experiment, as  $^{294}$Og  is known to  decay by
$\alpha$-decay with a half-life of 0.58\,ms \cite{Brewer2018}. This observation
could in fact also apply to the D1S results, since the D1S calculations  were
performed in a smaller collective space leading to overestimation of  the
half-lives \cite{Giuliani2014,Sadhukhan2014}.  It is to be noted that while
half-lives are very sensitive to details of the calculations,  the models used
in this study are very consistent with each other and with experiment when it
comes to global observables, such as alpha-decay energies, deformations, and
radii \cite{Heenen2015,Giuliani2019}. As demonstrated below,
spontaneous-fission mass and charge yields are also robustly predicted.

\begin{figure}[htb] 
 \includegraphics[width=1.0\linewidth]{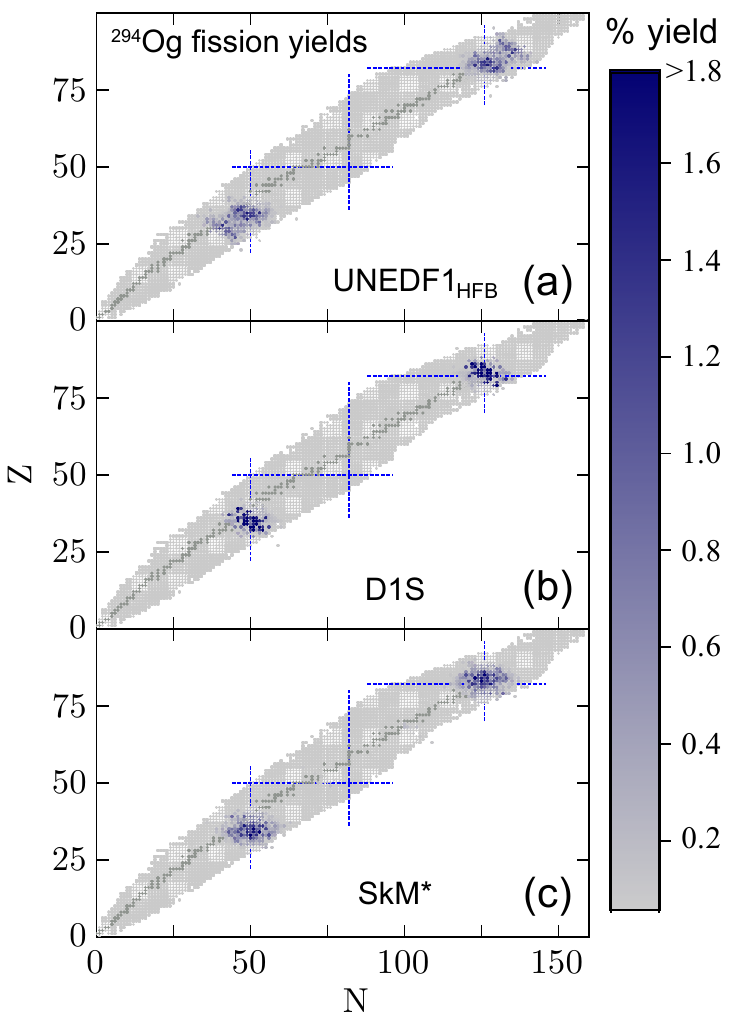}
	\caption[\Og{} N-Z fragment yields]{
	Fission fragment distributions for \Og{}  obtained in  \hfb{} (a), D1S
	(b), and SkM* (c) EDFs using the non-perturbative cranking ATDHFB
	inertia and  the baseline  dissipation tensor $\gras{\eta}_0$. Known
	isotopes are marked in grey \cite{NUDAT}. Magic numbers 50, 82, and 126
	are indicated by dotted lines.} \label{fig:2d-yield_atdhfb-np} 
\end{figure}

Despite the strong variations in the predicted $\tau_\textrm{SF}$, we see in
Fig.~\ref{fig:2d-yield_atdhfb-np} that the predicted fission yields are in fact
rather independent of the EDF choice. Namely,  all three functionals predict a
heavy fragment in the neighborhood of \Pb{} and a light fragment near \Kr{}. 

\begin{figure}[htb]  
\includegraphics[width=0.9\linewidth]{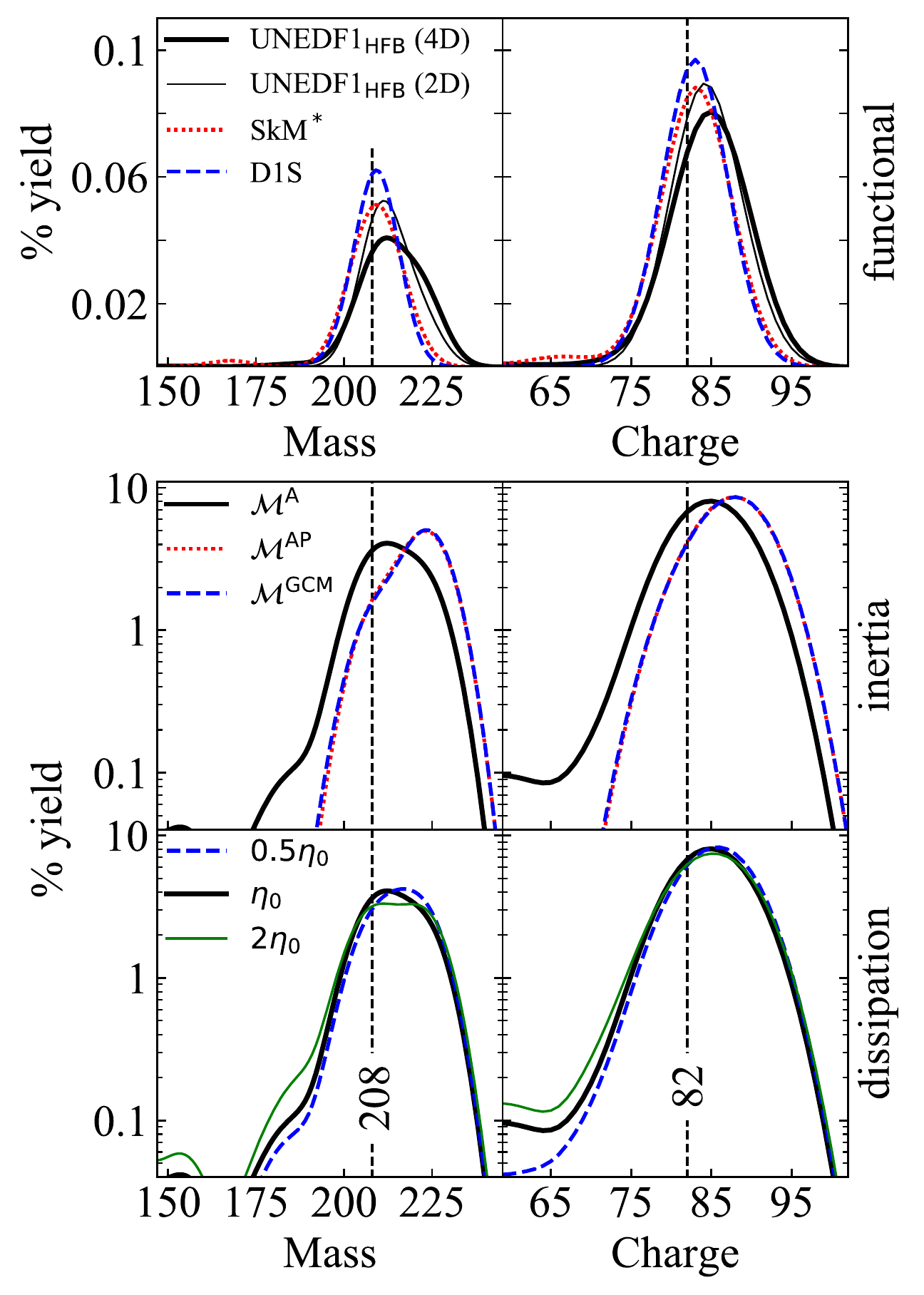}
	\caption[Comparison of \Og{} heavy fragment masses and
	charges]{Upper panel: Predicted heavy fragment mass (left) and charge
	(right) yields of \Og{} using different functionals (top, linear scale).
	Bottom panels: collective inertias  and dissipation tensor strengths (in
	logarithmic scale). The baseline calculation was performed using the
	\hfb{} functional in a 4D space with non-perturbative cranking ATDHFB
	inertia and dissipation tensor strength $\gras{\eta}_0$.} 
	\label{fig:comparisons} 
\end{figure}

The sensitivity to the different inputs are shown in Fig.~\ref{fig:comparisons}
through one-dimensional projections onto the fragment mass and charge. The top
panels of Fig.~\ref{fig:comparisons} shows again that all three functionals
predict highly-asymmetric fission with the heavy fragment centered at or around
\Pb{}. While the peaks corresponding to the D1S and  SkM* functionals overlap
quite well, the \hfb{} distributions (both in 2D and 4D space) are broader and
shifted slightly towards more asymmetric splits. This may be related to the
relative flatness of the \hfb{} PES compared to the others, which makes it more
susceptible to large fluctuations.  The secondary tiny peak around \Xe{}
predicted by SkM*, associated with the more symmetric fission valley  of
Fig.~\ref{fig:pes}(c) is clearly seen. For D1S and \hfb{}, the yield
distributions do not show a tail at lower masses/charges. As discussed below, 
this can be associated with both the
collective inertia  and the energy ridge (particularly pronounced for D1S), both
effectively separating  the two fission valleys.


The impact of the choice of collective inertia on the fission yield
distributions is illustrated in the middle panels of Fig.~\ref{fig:comparisons}.
While recent time-dependent DFT work on induced fission has played down the role
of collective inertia \cite{Bulgac2018} outside the barrier, it was emphasized
in Ref.~\cite{Sadhukhan2016} that the tunneling phase of spontaneous fission was
highly sensitive to it. This may explain the fine deviations observed in
Fig.~\ref{fig:comparisons}. The yields corresponding to {\MATDHFp} and {\MGCMp}
overlap and both are shifted towards more asymmetric splits compared to
{\MATDHF}. This suggests that it is the topology of the collective inertia,
rather than its absolute magnitude, which affects the shape of the fission
yields. In particular, we find that the smoothness of the perturbative inertia
allows fluctuations to diffuse the yield to more extreme fragment
configurations.  

The bottom panels of Fig.~\ref{fig:comparisons} show the effect of varying the
strength of the dissipation tensor. This parameter  has a noticeable impact on
the yields, particularly on the tails and the yields associated with the
more-symmetric channel.  The results corresponding to
$\gras{\eta}=\gras{\eta}_0,0.5\gras{\eta}_0$, and $2\gras{\eta}_0$ are very
close. This is consistent with findings of Refs.
\cite{Randrup2011,Sierk2017,Sadhukhan2017}, which found that the yield
distributions are not very sensitive to the precise value of the dissipation
tensor.  In general, irrespective of the choice of energy density functional, we
found a rather similar pattern of yield distributions with respect to the
inertia tensor and the dissipation strength. 
Only at the extreme limits of constant inertia and no dissipation do we see 
an increase of the symmetric peak in the predicted yields. However, as discussed in
\cite{Sadhukhan2013,Sadhukhan2014,Arimoto2014,Sierk2017,Sadhukhan2017}, such 
choices are not realistic. For this reason, we have not included them in the figure and mention 
them only for the sake of completeness.

\begin{figure}[htb]
	\includegraphics[width=0.95\linewidth]{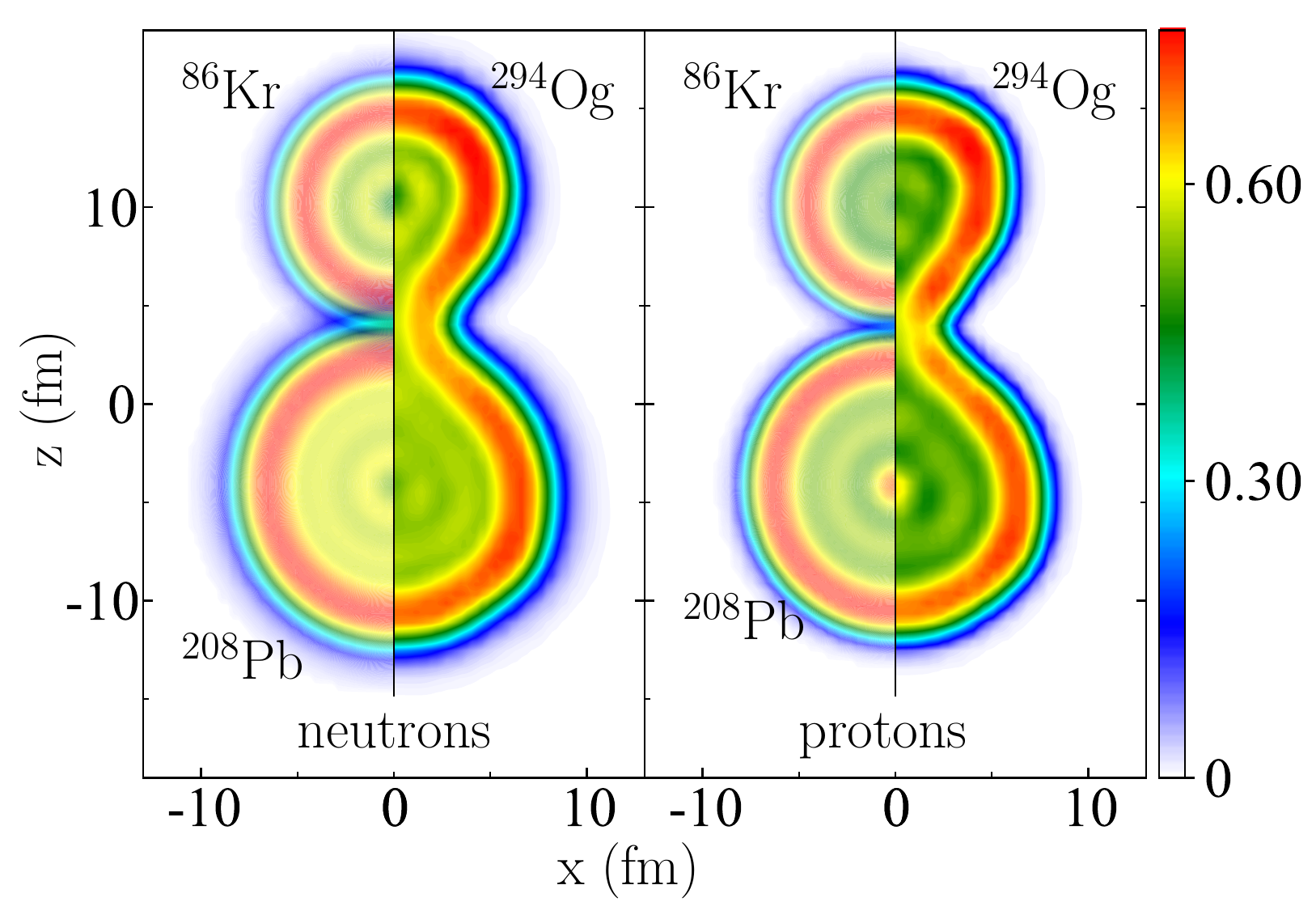}
	\caption[]{Nucleon localization function for a highly-deformed
	configuration of \Og, $(Q_{20},Q_{30})=(264 \, \mathrm{b}, \, 60
	\,\mathrm{b}^{3/2})$ for neutrons (left panels) and protons (right
	panels). For comparison, localizations are shown for the prefragments {\Pb}
	and {\Kr} on the left side of each
	subplot.\label{fig:og294-withfragssplitscreen}
	}
\end{figure}

Finally, to better understand the formation of the most-probable fission
fragments, we studied the nucleon localization functions
\cite{Zhang2016,Sadhukhan2017} along the cluster-decay path.  As shown in
Fig.~\ref{fig:og294-withfragssplitscreen}, by comparing the \Og{} localizations
with those calculated for spherical \Pb{} and \Kr{}, we found that both the lead prefragment
and the $N\approx50$ neutrons belonging to krypton are well-localized. This
result highlights the importance of shell structure along the fission path in
determining the most probable fragment configuration in fissioning nuclei. It is
interesting to note that the neutron localizations in the highly elongated
configurations of {\Og} shown in the left panels of
Fig.~\ref{fig:og294-withfragssplitscreen} have much more structure than the
ground-state localizations, which are close to the Fermi-gas limit
\cite{Jerabek2018}. Indeed, the single-particle level density is very high in {\Og}, which   is not the case for the prefragments.

{\it Conclusions} -- In this paper, which reinforces the results of
Ref.~\cite{Warda2018}, we predict that the dominant spontaneous fission mode of
\Og{} will be a highly-asymmetric cluster emission sharply centered around
doubly-magic  \Pb{} and magic \Kr{}. We have shown that this prediction is
fairly  robust with respect to the choice of input parameters, such as  energy
density functional, collective inertia, and dissipation tensor. In particular,
we emphasize that differences in barrier heights predicted by different EDFs do
not affect the calculated fission yields.  We confirmed the implicit assumption
of \cite{Sadhukhan2016}, that 4D calculations do not necessarily offer an
improved description of the tunneling compared to a well-chosen 3D description,
and we argue for a hierarchy of ingredients necessary for a Langevin description
of low-energy fission.

The search for cluster emission has already begun \cite{Brewer2018}. Since, as
demonstrated in this work, an asymmetric fission of \Og{} is expected to be
strongly enhanced thus providing a strong experimental trigger, we are
optimistic that the next generation of experiments, perhaps involving an
ionization chamber \cite{Brewer2018},  will find experimental evidence of
cluster radioactivity of \Og{}.

\section{Acknowledgements}\label{sec:acknowledgements}

This work was  supported by the U.S. Department of Energy under Award Numbers
DOE-DE-NA0002847 (NNSA, the Stewardship Science Academic Alliances program),
DE-SC0013365 (Office of Science), and DE-SC0018083 (Office of Science, NUCLEI
SciDAC-4 collaboration). as well as by the Office of Science, Office of
Workforce Development for Teachers and Scientists, Office of Science Graduate
Student Research (SCGSR) program. The SCGSR program is administered by the Oak
Ridge Institute for Science and Education (ORISE) for the DOE\@. ORISE is managed
by ORAU under contract number DE-SC0014664. This work was also partly performed
under the auspices of the US Department of Energy by the Lawrence Livermore
National Laboratory under Contract DE-AC52-07NA27344. Computing support came
from the Lawrence Livermore National Laboratory (LLNL) Institutional Computing
Grand Challenge program.

\bibliographystyle{apsrev4-1}
\bibliography{main}

\begin{thebibliography}{61}%
\makeatletter
\providecommand \@ifxundefined [1]{%
 \@ifx{#1\undefined}
}%
\providecommand \@ifnum [1]{%
 \ifnum #1\expandafter \@firstoftwo
 \else \expandafter \@secondoftwo
 \fi
}%
\providecommand \@ifx [1]{%
 \ifx #1\expandafter \@firstoftwo
 \else \expandafter \@secondoftwo
 \fi
}%
\providecommand \natexlab [1]{#1}%
\providecommand \enquote  [1]{``#1''}%
\providecommand \bibnamefont  [1]{#1}%
\providecommand \bibfnamefont [1]{#1}%
\providecommand \citenamefont [1]{#1}%
\providecommand \href@noop [0]{\@secondoftwo}%
\providecommand \href [0]{\begingroup \@sanitize@url \@href}%
\providecommand \@href[1]{\@@startlink{#1}\@@href}%
\providecommand \@@href[1]{\endgroup#1\@@endlink}%
\providecommand \@sanitize@url [0]{\catcode `\\12\catcode `\$12\catcode
  `\&12\catcode `\#12\catcode `\^12\catcode `\_12\catcode `\%12\relax}%
\providecommand \@@startlink[1]{}%
\providecommand \@@endlink[0]{}%
\providecommand \url  [0]{\begingroup\@sanitize@url \@url }%
\providecommand \@url [1]{\endgroup\@href {#1}{\urlprefix }}%
\providecommand \urlprefix  [0]{URL }%
\providecommand \Eprint [0]{\href }%
\providecommand \doibase [0]{http://dx.doi.org/}%
\providecommand \selectlanguage [0]{\@gobble}%
\providecommand \bibinfo  [0]{\@secondoftwo}%
\providecommand \bibfield  [0]{\@secondoftwo}%
\providecommand \translation [1]{[#1]}%
\providecommand \BibitemOpen [0]{}%
\providecommand \bibitemStop [0]{}%
\providecommand \bibitemNoStop [0]{.\EOS\space}%
\providecommand \EOS [0]{\spacefactor3000\relax}%
\providecommand \BibitemShut  [1]{\csname bibitem#1\endcsname}%
\let\auto@bib@innerbib\@empty
\bibitem [{\citenamefont {Oganessian}\ \emph {et~al.}(2006)\citenamefont
  {Oganessian} \emph {et~al.}}]{Oganessian2006}%
  \BibitemOpen
  \bibfield  {author} {\bibinfo {author} {\bibfnamefont {Y.~T.}\ \bibnamefont
  {Oganessian}} \emph {et~al.},\ }\href {\doibase 10.1103/PhysRevC.74.044602}
  {\bibfield  {journal} {\bibinfo  {journal} {Phys. Rev. C}\ }\textbf {\bibinfo
  {volume} {74}},\ \bibinfo {pages} {044602} (\bibinfo {year}
  {2006})}\BibitemShut {NoStop}%
\bibitem [{\citenamefont {Oganessian}\ \emph {et~al.}(2012)\citenamefont
  {Oganessian} \emph {et~al.}}]{Oganessian2012}%
  \BibitemOpen
  \bibfield  {author} {\bibinfo {author} {\bibfnamefont {Y.~T.}\ \bibnamefont
  {Oganessian}} \emph {et~al.},\ }\href {\doibase
  10.1103/PhysRevLett.109.162501} {\bibfield  {journal} {\bibinfo  {journal}
  {Phys. Rev. Lett.}\ }\textbf {\bibinfo {volume} {109}},\ \bibinfo {pages}
  {162501} (\bibinfo {year} {2012})}\BibitemShut {NoStop}%
\bibitem [{\citenamefont {Brewer}\ \emph {et~al.}(2018)\citenamefont {Brewer}
  \emph {et~al.}}]{Brewer2018}%
  \BibitemOpen
  \bibfield  {author} {\bibinfo {author} {\bibfnamefont {N.}~\bibnamefont
  {Brewer}} \emph {et~al.},\ }\href {\doibase 10.1103/PhysRevC.98.024317}
  {\bibfield  {journal} {\bibinfo  {journal} {Phys. Rev. C}\ }\textbf {\bibinfo
  {volume} {98}},\ \bibinfo {pages} {024317} (\bibinfo {year}
  {2018})}\BibitemShut {NoStop}%
\bibitem [{\citenamefont {Dmitriev}\ \emph {et~al.}(2016)\citenamefont
  {Dmitriev}, \citenamefont {Itkis},\ and\ \citenamefont
  {Oganessian}}]{Dmitriev2016}%
  \BibitemOpen
  \bibfield  {author} {\bibinfo {author} {\bibfnamefont {S.}~\bibnamefont
  {Dmitriev}}, \bibinfo {author} {\bibfnamefont {M.}~\bibnamefont {Itkis}}, \
  and\ \bibinfo {author} {\bibfnamefont {Y.}~\bibnamefont {Oganessian}},\
  }\href {\doibase 10.1051/epjconf/201613108001} {\bibfield  {journal}
  {\bibinfo  {journal} {Eur. Phys. J. WOC}\ }\textbf {\bibinfo {volume}
  {131}},\ \bibinfo {pages} {08001} (\bibinfo {year} {2016})}\BibitemShut
  {NoStop}%
\bibitem [{\citenamefont {Oganessian}\ and\ \citenamefont
  {Dmitriev}(2016)}]{Oganessian2016}%
  \BibitemOpen
  \bibfield  {author} {\bibinfo {author} {\bibfnamefont {Y.~T.}\ \bibnamefont
  {Oganessian}}\ and\ \bibinfo {author} {\bibfnamefont {S.~N.}\ \bibnamefont
  {Dmitriev}},\ }\href {http://stacks.iop.org/0036-021X/85/i=9/a=901}
  {\bibfield  {journal} {\bibinfo  {journal} {Russ. Chem. Rev.}\ }\textbf
  {\bibinfo {volume} {85}},\ \bibinfo {pages} {901} (\bibinfo {year}
  {2016})}\BibitemShut {NoStop}%
\bibitem [{\citenamefont {Hofmann}\ \emph {et~al.}(2016)\citenamefont {Hofmann}
  \emph {et~al.}}]{Hoffman2016}%
  \BibitemOpen
  \bibfield  {author} {\bibinfo {author} {\bibfnamefont {S.}~\bibnamefont
  {Hofmann}} \emph {et~al.},\ }\href {\doibase 10.1140/epja/i2016-16180-4}
  {\bibfield  {journal} {\bibinfo  {journal} {Eur. Phys. J. A}\ }\textbf
  {\bibinfo {volume} {52}},\ \bibinfo {pages} {180} (\bibinfo {year}
  {2016})}\BibitemShut {NoStop}%
\bibitem [{\citenamefont {Roberto}\ and\ \citenamefont
  {Rykaczewski}(2018)}]{Roberto2018}%
  \BibitemOpen
  \bibfield  {author} {\bibinfo {author} {\bibfnamefont {J.~B.}\ \bibnamefont
  {Roberto}}\ and\ \bibinfo {author} {\bibfnamefont {K.~P.}\ \bibnamefont
  {Rykaczewski}},\ }\href {\doibase 10.1080/01496395.2017.1290658} {\bibfield
  {journal} {\bibinfo  {journal} {Sep. Sci. Technol.}\ }\textbf {\bibinfo
  {volume} {53}},\ \bibinfo {pages} {1813} (\bibinfo {year}
  {2018})}\BibitemShut {NoStop}%
\bibitem [{\citenamefont {D{\"u}llmann}\ and\ \citenamefont
  {Block}(2018)}]{Duellmann2018}%
  \BibitemOpen
  \bibfield  {author} {\bibinfo {author} {\bibfnamefont {C.~E.}\ \bibnamefont
  {D{\"u}llmann}}\ and\ \bibinfo {author} {\bibfnamefont {M.}~\bibnamefont
  {Block}},\ }\href
  {https://www.scientificamerican.com/article/the-quest-for-superheavy-elements-and-the-island-of-stability/}
  {\bibfield  {journal} {\bibinfo  {journal} {Sci. Am.}\ }\textbf {\bibinfo
  {volume} {318}},\ \bibinfo {pages} {48} (\bibinfo {year} {2018})}\BibitemShut
  {NoStop}%
\bibitem [{\citenamefont {Jerabek}\ \emph {et~al.}(2018)\citenamefont
  {Jerabek}, \citenamefont {Schuetrumpf}, \citenamefont {Schwerdtfeger},\ and\
  \citenamefont {Nazarewicz}}]{Jerabek2018}%
  \BibitemOpen
  \bibfield  {author} {\bibinfo {author} {\bibfnamefont {P.}~\bibnamefont
  {Jerabek}}, \bibinfo {author} {\bibfnamefont {B.}~\bibnamefont
  {Schuetrumpf}}, \bibinfo {author} {\bibfnamefont {P.}~\bibnamefont
  {Schwerdtfeger}}, \ and\ \bibinfo {author} {\bibfnamefont {W.}~\bibnamefont
  {Nazarewicz}},\ }\href {\doibase 10.1103/PhysRevLett.120.053001} {\bibfield
  {journal} {\bibinfo  {journal} {Phys. Rev. Lett.}\ }\textbf {\bibinfo
  {volume} {120}},\ \bibinfo {pages} {5} (\bibinfo {year} {2018})}\BibitemShut
  {NoStop}%
\bibitem [{\citenamefont {Nazarewicz}(2018)}]{Nazarewicz2018}%
  \BibitemOpen
  \bibfield  {author} {\bibinfo {author} {\bibfnamefont {W.}~\bibnamefont
  {Nazarewicz}},\ }\href {\doibase 10.1038/s41567-018-0163-3} {\bibfield
  {journal} {\bibinfo  {journal} {Nature Phys.}\ }\textbf {\bibinfo {volume}
  {14}},\ \bibinfo {pages} {537} (\bibinfo {year} {2018})}\BibitemShut
  {NoStop}%
\bibitem [{\citenamefont {Giuliani}\ \emph {et~al.}(2019)\citenamefont
  {Giuliani}, \citenamefont {Matheson}, \citenamefont {Nazarewicz},
  \citenamefont {Olsen}, \citenamefont {Reinhard}, \citenamefont {Sadhukhan},
  \citenamefont {Schuetrumpf}, \citenamefont {Schunck},\ and\ \citenamefont
  {Schwerdtfeger}}]{Giuliani2019}%
  \BibitemOpen
  \bibfield  {author} {\bibinfo {author} {\bibfnamefont {S.~A.}\ \bibnamefont
  {Giuliani}}, \bibinfo {author} {\bibfnamefont {Z.}~\bibnamefont {Matheson}},
  \bibinfo {author} {\bibfnamefont {W.}~\bibnamefont {Nazarewicz}}, \bibinfo
  {author} {\bibfnamefont {E.}~\bibnamefont {Olsen}}, \bibinfo {author}
  {\bibfnamefont {P.-G.}\ \bibnamefont {Reinhard}}, \bibinfo {author}
  {\bibfnamefont {J.}~\bibnamefont {Sadhukhan}}, \bibinfo {author}
  {\bibfnamefont {B.}~\bibnamefont {Schuetrumpf}}, \bibinfo {author}
  {\bibfnamefont {N.}~\bibnamefont {Schunck}}, \ and\ \bibinfo {author}
  {\bibfnamefont {P.}~\bibnamefont {Schwerdtfeger}},\ }\href@noop {} {\bibfield
   {journal} {\bibinfo  {journal} {Rev. Mod. Phys., in press}\ } (\bibinfo
  {year} {2019})}\BibitemShut {NoStop}%
\bibitem [{\citenamefont {He{\ss}berger}(2017)}]{Hessberger2017}%
  \BibitemOpen
  \bibfield  {author} {\bibinfo {author} {\bibfnamefont {F.~P.}\ \bibnamefont
  {He{\ss}berger}},\ }\href {\doibase 10.1140/epja/i2017-12260-3} {\bibfield
  {journal} {\bibinfo  {journal} {Euro. Phys. J. A}\ }\textbf {\bibinfo
  {volume} {53}},\ \bibinfo {pages} {75} (\bibinfo {year} {2017})}\BibitemShut
  {NoStop}%
\bibitem [{\citenamefont {Baran}\ \emph {et~al.}(2015)\citenamefont {Baran},
  \citenamefont {Kowal}, \citenamefont {Reinhard}, \citenamefont {Robledo},
  \citenamefont {Staszczak},\ and\ \citenamefont {Warda}}]{Baran2015}%
  \BibitemOpen
  \bibfield  {author} {\bibinfo {author} {\bibfnamefont {A.}~\bibnamefont
  {Baran}}, \bibinfo {author} {\bibfnamefont {M.}~\bibnamefont {Kowal}},
  \bibinfo {author} {\bibfnamefont {P.~G.}\ \bibnamefont {Reinhard}}, \bibinfo
  {author} {\bibfnamefont {L.~M.}\ \bibnamefont {Robledo}}, \bibinfo {author}
  {\bibfnamefont {A.}~\bibnamefont {Staszczak}}, \ and\ \bibinfo {author}
  {\bibfnamefont {M.}~\bibnamefont {Warda}},\ }\href {\doibase
  10.1016/j.nuclphysa.2015.06.002} {\bibfield  {journal} {\bibinfo  {journal}
  {Nucl. Phys. A}\ }\textbf {\bibinfo {volume} {944}},\ \bibinfo {pages} {442}
  (\bibinfo {year} {2015})}\BibitemShut {NoStop}%
\bibitem [{\citenamefont {Giuliani}\ \emph {et~al.}(2018)\citenamefont
  {Giuliani}, \citenamefont {Mart{\'{i}}nez-Pinedo},\ and\ \citenamefont
  {Robledo}}]{Giuliani2018}%
  \BibitemOpen
  \bibfield  {author} {\bibinfo {author} {\bibfnamefont {S.~A.}\ \bibnamefont
  {Giuliani}}, \bibinfo {author} {\bibfnamefont {G.}~\bibnamefont
  {Mart{\'{i}}nez-Pinedo}}, \ and\ \bibinfo {author} {\bibfnamefont {L.~M.}\
  \bibnamefont {Robledo}},\ }\href {\doibase 10.1103/PhysRevC.97.034323}
  {\bibfield  {journal} {\bibinfo  {journal} {Phys. Rev. C}\ }\textbf {\bibinfo
  {volume} {97}},\ \bibinfo {pages} {034323} (\bibinfo {year}
  {2018})}\BibitemShut {NoStop}%
\bibitem [{\citenamefont {Schunck}\ and\ \citenamefont
  {Robledo}(2016)}]{Schunck2016}%
  \BibitemOpen
  \bibfield  {author} {\bibinfo {author} {\bibfnamefont {N.}~\bibnamefont
  {Schunck}}\ and\ \bibinfo {author} {\bibfnamefont {L.~M.}\ \bibnamefont
  {Robledo}},\ }\href {\doibase 10.1088/0034-4885/79/11/116301} {\bibfield
  {journal} {\bibinfo  {journal} {Rep. Prog. Phys.}\ }\textbf {\bibinfo
  {volume} {79}},\ \bibinfo {pages} {116301} (\bibinfo {year}
  {2016})}\BibitemShut {NoStop}%
\bibitem [{\citenamefont {Erler}\ \emph {et~al.}(2012)\citenamefont {Erler},
  \citenamefont {Langanke}, \citenamefont {Loens}, \citenamefont
  {Mart{\'{i}}nez-Pinedo},\ and\ \citenamefont {Reinhard}}]{Erler2012}%
  \BibitemOpen
  \bibfield  {author} {\bibinfo {author} {\bibfnamefont {J.}~\bibnamefont
  {Erler}}, \bibinfo {author} {\bibfnamefont {K.}~\bibnamefont {Langanke}},
  \bibinfo {author} {\bibfnamefont {H.~P.}\ \bibnamefont {Loens}}, \bibinfo
  {author} {\bibfnamefont {G.}~\bibnamefont {Mart{\'{i}}nez-Pinedo}}, \ and\
  \bibinfo {author} {\bibfnamefont {P.-G.}\ \bibnamefont {Reinhard}},\ }\href
  {\doibase 10.1103/PhysRevC.85.025802} {\bibfield  {journal} {\bibinfo
  {journal} {Phys. Rev. C}\ }\textbf {\bibinfo {volume} {85}},\ \bibinfo
  {pages} {025802} (\bibinfo {year} {2012})}\BibitemShut {NoStop}%
\bibitem [{\citenamefont {Staszczak}\ \emph {et~al.}(2013)\citenamefont
  {Staszczak}, \citenamefont {Baran},\ and\ \citenamefont
  {Nazarewicz}}]{Staszczak2013}%
  \BibitemOpen
  \bibfield  {author} {\bibinfo {author} {\bibfnamefont {A.}~\bibnamefont
  {Staszczak}}, \bibinfo {author} {\bibfnamefont {A.}~\bibnamefont {Baran}}, \
  and\ \bibinfo {author} {\bibfnamefont {W.}~\bibnamefont {Nazarewicz}},\
  }\href {\doibase 10.1103/PhysRevC.87.024320} {\bibfield  {journal} {\bibinfo
  {journal} {Phys. Rev. C}\ }\textbf {\bibinfo {volume} {87}},\ \bibinfo
  {pages} {024320} (\bibinfo {year} {2013})}\BibitemShut {NoStop}%
\bibitem [{\citenamefont {Giuliani}\ and\ \citenamefont
  {Robledo}(2013)}]{Giuliani2013}%
  \BibitemOpen
  \bibfield  {author} {\bibinfo {author} {\bibfnamefont {S.~A.}\ \bibnamefont
  {Giuliani}}\ and\ \bibinfo {author} {\bibfnamefont {L.~M.}\ \bibnamefont
  {Robledo}},\ }\href {\doibase 10.1103/PhysRevC.88.054325} {\bibfield
  {journal} {\bibinfo  {journal} {Phys. Rev. C}\ }\textbf {\bibinfo {volume}
  {88}},\ \bibinfo {pages} {054325} (\bibinfo {year} {2013})}\BibitemShut
  {NoStop}%
\bibitem [{\citenamefont {Giuliani}\ \emph {et~al.}(2014)\citenamefont
  {Giuliani}, \citenamefont {Robledo},\ and\ \citenamefont
  {Rodr\'{\i}guez-Guzm\'an}}]{Giuliani2014}%
  \BibitemOpen
  \bibfield  {author} {\bibinfo {author} {\bibfnamefont {S.~A.}\ \bibnamefont
  {Giuliani}}, \bibinfo {author} {\bibfnamefont {L.~M.}\ \bibnamefont
  {Robledo}}, \ and\ \bibinfo {author} {\bibfnamefont {R.}~\bibnamefont
  {Rodr\'{\i}guez-Guzm\'an}},\ }\href {\doibase 10.1103/PhysRevC.90.054311}
  {\bibfield  {journal} {\bibinfo  {journal} {Phys. Rev. C}\ }\textbf {\bibinfo
  {volume} {90}},\ \bibinfo {pages} {054311} (\bibinfo {year}
  {2014})}\BibitemShut {NoStop}%
\bibitem [{\citenamefont {Lema{\^{i}}tre}\ \emph {et~al.}(2018)\citenamefont
  {Lema{\^{i}}tre}, \citenamefont {Goriely}, \citenamefont {Hilaire},\ and\
  \citenamefont {Dubray}}]{Lemaitre2018}%
  \BibitemOpen
  \bibfield  {author} {\bibinfo {author} {\bibfnamefont {J.-F.}\ \bibnamefont
  {Lema{\^{i}}tre}}, \bibinfo {author} {\bibfnamefont {S.}~\bibnamefont
  {Goriely}}, \bibinfo {author} {\bibfnamefont {S.}~\bibnamefont {Hilaire}}, \
  and\ \bibinfo {author} {\bibfnamefont {N.}~\bibnamefont {Dubray}},\ }\href
  {\doibase 10.1103/PhysRevC.98.024623} {\bibfield  {journal} {\bibinfo
  {journal} {Phys. Rev. C}\ }\textbf {\bibinfo {volume} {98}},\ \bibinfo
  {pages} {024623} (\bibinfo {year} {2018})}\BibitemShut {NoStop}%
\bibitem [{\citenamefont {Rodr\'{\i}guez-Guzm\'an}\ and\ \citenamefont
  {Robledo}(2018)}]{Rodriguez2018}%
  \BibitemOpen
  \bibfield  {author} {\bibinfo {author} {\bibfnamefont {R.}~\bibnamefont
  {Rodr\'{\i}guez-Guzm\'an}}\ and\ \bibinfo {author} {\bibfnamefont {L.~M.}\
  \bibnamefont {Robledo}},\ }\href {\doibase 10.1103/PhysRevC.98.034308}
  {\bibfield  {journal} {\bibinfo  {journal} {Phys. Rev. C}\ }\textbf {\bibinfo
  {volume} {98}},\ \bibinfo {pages} {034308} (\bibinfo {year}
  {2018})}\BibitemShut {NoStop}%
\bibitem [{\citenamefont {Sadhukhan}\ \emph {et~al.}(2016)\citenamefont
  {Sadhukhan}, \citenamefont {Nazarewicz},\ and\ \citenamefont
  {Schunck}}]{Sadhukhan2016}%
  \BibitemOpen
  \bibfield  {author} {\bibinfo {author} {\bibfnamefont {J.}~\bibnamefont
  {Sadhukhan}}, \bibinfo {author} {\bibfnamefont {W.}~\bibnamefont
  {Nazarewicz}}, \ and\ \bibinfo {author} {\bibfnamefont {N.}~\bibnamefont
  {Schunck}},\ }\href {\doibase 10.1103/PhysRevC.93.011304} {\bibfield
  {journal} {\bibinfo  {journal} {Phys. Rev. C}\ }\textbf {\bibinfo {volume}
  {93}},\ \bibinfo {pages} {011304} (\bibinfo {year} {2016})}\BibitemShut
  {NoStop}%
\bibitem [{\citenamefont {Warda}\ \emph {et~al.}(2018)\citenamefont {Warda},
  \citenamefont {Zdeb},\ and\ \citenamefont {Robledo}}]{Warda2018}%
  \BibitemOpen
  \bibfield  {author} {\bibinfo {author} {\bibfnamefont {M.}~\bibnamefont
  {Warda}}, \bibinfo {author} {\bibfnamefont {A.}~\bibnamefont {Zdeb}}, \ and\
  \bibinfo {author} {\bibfnamefont {L.~M.}\ \bibnamefont {Robledo}},\ }\href
  {\doibase 10.1103/PhysRevC.98.041602} {\bibfield  {journal} {\bibinfo
  {journal} {Phys. Rev. C}\ }\textbf {\bibinfo {volume} {98}},\ \bibinfo
  {pages} {041602} (\bibinfo {year} {2018})}\BibitemShut {NoStop}%
\bibitem [{\citenamefont {Regnier}\ \emph {et~al.}(2016)\citenamefont
  {Regnier}, \citenamefont {Dubray}, \citenamefont {Schunck},\ and\
  \citenamefont {Verri\`ere}}]{Regnier2016}%
  \BibitemOpen
  \bibfield  {author} {\bibinfo {author} {\bibfnamefont {D.}~\bibnamefont
  {Regnier}}, \bibinfo {author} {\bibfnamefont {N.}~\bibnamefont {Dubray}},
  \bibinfo {author} {\bibfnamefont {N.}~\bibnamefont {Schunck}}, \ and\
  \bibinfo {author} {\bibfnamefont {M.}~\bibnamefont {Verri\`ere}},\ }\href
  {\doibase 10.1103/PhysRevC.93.054611} {\bibfield  {journal} {\bibinfo
  {journal} {Phys. Rev. C}\ }\textbf {\bibinfo {volume} {93}},\ \bibinfo
  {pages} {054611} (\bibinfo {year} {2016})}\BibitemShut {NoStop}%
\bibitem [{\citenamefont {Regnier}\ \emph {et~al.}(2019)\citenamefont
  {Regnier}, \citenamefont {Dubray},\ and\ \citenamefont
  {Schunck}}]{Regnier2018}%
  \BibitemOpen
  \bibfield  {author} {\bibinfo {author} {\bibfnamefont {D.}~\bibnamefont
  {Regnier}}, \bibinfo {author} {\bibfnamefont {N.}~\bibnamefont {Dubray}}, \
  and\ \bibinfo {author} {\bibfnamefont {N.}~\bibnamefont {Schunck}},\ }\href
  {\doibase 10.1103/PhysRevC.99.024611} {\bibfield  {journal} {\bibinfo
  {journal} {Phys. Rev. C}\ }\textbf {\bibinfo {volume} {99}},\ \bibinfo
  {pages} {024611} (\bibinfo {year} {2019})}\BibitemShut {NoStop}%
\bibitem [{\citenamefont {Sandulescu}\ \emph {et~al.}(1980)\citenamefont
  {Sandulescu}, \citenamefont {Poenaru},\ and\ \citenamefont
  {Greiner}}]{Sandulescu1980}%
  \BibitemOpen
  \bibfield  {author} {\bibinfo {author} {\bibfnamefont {A.}~\bibnamefont
  {Sandulescu}}, \bibinfo {author} {\bibfnamefont {D.}~\bibnamefont {Poenaru}},
  \ and\ \bibinfo {author} {\bibfnamefont {W.}~\bibnamefont {Greiner}},\
  }\href@noop {} {\bibfield  {journal} {\bibinfo  {journal} {Sov. J. Part.
  Nuclei}\ }\textbf {\bibinfo {volume} {11}},\ \bibinfo {pages} {528} (\bibinfo
  {year} {1980})}\BibitemShut {NoStop}%
\bibitem [{\citenamefont {Poenaru}\ \emph {et~al.}(1986)\citenamefont
  {Poenaru}, \citenamefont {Greiner}, \citenamefont {Depta}, \citenamefont
  {Ivascu}, \citenamefont {Mazilu},\ and\ \citenamefont
  {Sandulescu}}]{Poenaru1986}%
  \BibitemOpen
  \bibfield  {author} {\bibinfo {author} {\bibfnamefont {D.}~\bibnamefont
  {Poenaru}}, \bibinfo {author} {\bibfnamefont {W.}~\bibnamefont {Greiner}},
  \bibinfo {author} {\bibfnamefont {K.}~\bibnamefont {Depta}}, \bibinfo
  {author} {\bibfnamefont {M.}~\bibnamefont {Ivascu}}, \bibinfo {author}
  {\bibfnamefont {D.}~\bibnamefont {Mazilu}}, \ and\ \bibinfo {author}
  {\bibfnamefont {A.}~\bibnamefont {Sandulescu}},\ }\href {\doibase
  https://doi.org/10.1016/0092-640X(86)90013-6} {\bibfield  {journal} {\bibinfo
   {journal} {At. Data Nucl. Data Tables}\ }\textbf {\bibinfo {volume} {34}},\
  \bibinfo {pages} {423 } (\bibinfo {year} {1986})}\BibitemShut {NoStop}%
\bibitem [{\citenamefont {Royer}\ \emph {et~al.}(1998)\citenamefont {Royer},
  \citenamefont {Gupta},\ and\ \citenamefont {Denisov}}]{Royer1998}%
  \BibitemOpen
  \bibfield  {author} {\bibinfo {author} {\bibfnamefont {G.}~\bibnamefont
  {Royer}}, \bibinfo {author} {\bibfnamefont {R.~K.}\ \bibnamefont {Gupta}}, \
  and\ \bibinfo {author} {\bibfnamefont {V.}~\bibnamefont {Denisov}},\ }\href
  {\doibase https://doi.org/10.1016/S0375-9474(97)00801-4} {\bibfield
  {journal} {\bibinfo  {journal} {Nucl. Phys. A}\ }\textbf {\bibinfo {volume}
  {632}},\ \bibinfo {pages} {275 } (\bibinfo {year} {1998})}\BibitemShut
  {NoStop}%
\bibitem [{\citenamefont {Rose}\ and\ \citenamefont {Jones}(1983)}]{Rose1983}%
  \BibitemOpen
  \bibfield  {author} {\bibinfo {author} {\bibfnamefont {H.~J.}\ \bibnamefont
  {Rose}}\ and\ \bibinfo {author} {\bibfnamefont {G.~A.}\ \bibnamefont
  {Jones}},\ }\href {https://doi.org/10.1038/307245a0} {\bibfield  {journal}
  {\bibinfo  {journal} {Nature}\ }\textbf {\bibinfo {volume} {307}},\ \bibinfo
  {pages} {245–247} (\bibinfo {year} {1983})}\BibitemShut {NoStop}%
\bibitem [{\citenamefont {Poenaru}\ and\ \citenamefont
  {Greiner}(2010)}]{Poenaru2010}%
  \BibitemOpen
  \bibfield  {author} {\bibinfo {author} {\bibfnamefont {D.~N.}\ \bibnamefont
  {Poenaru}}\ and\ \bibinfo {author} {\bibfnamefont {W.}~\bibnamefont
  {Greiner}},\ }in\ \href {\doibase 10.1007/978-3-642-13899-7_1} {\emph
  {\bibinfo {booktitle} {Clusters in Nuclei: Volume 1}}},\ \bibinfo {editor}
  {edited by\ \bibinfo {editor} {\bibfnamefont {C.}~\bibnamefont {Beck}}}\
  (\bibinfo  {publisher} {Springer Berlin Heidelberg},\ \bibinfo {address}
  {Berlin, Heidelberg},\ \bibinfo {year} {2010})\ p.~\bibinfo {pages}
  {1}\BibitemShut {NoStop}%
\bibitem [{\citenamefont {Poenaru}\ \emph {et~al.}(2011)\citenamefont
  {Poenaru}, \citenamefont {Gherghescu},\ and\ \citenamefont
  {Greiner}}]{Poenaru2011}%
  \BibitemOpen
  \bibfield  {author} {\bibinfo {author} {\bibfnamefont {D.~N.}\ \bibnamefont
  {Poenaru}}, \bibinfo {author} {\bibfnamefont {R.~A.}\ \bibnamefont
  {Gherghescu}}, \ and\ \bibinfo {author} {\bibfnamefont {W.}~\bibnamefont
  {Greiner}},\ }\href {\doibase 10.1103/PhysRevLett.107.062503} {\bibfield
  {journal} {\bibinfo  {journal} {Phys. Rev. Lett.}\ }\textbf {\bibinfo
  {volume} {107}},\ \bibinfo {pages} {062503} (\bibinfo {year}
  {2011})}\BibitemShut {NoStop}%
\bibitem [{\citenamefont {Poenaru}\ \emph {et~al.}(2012)\citenamefont
  {Poenaru}, \citenamefont {Gherghescu},\ and\ \citenamefont
  {Greiner}}]{Poenaru2012}%
  \BibitemOpen
  \bibfield  {author} {\bibinfo {author} {\bibfnamefont {D.~N.}\ \bibnamefont
  {Poenaru}}, \bibinfo {author} {\bibfnamefont {R.~A.}\ \bibnamefont
  {Gherghescu}}, \ and\ \bibinfo {author} {\bibfnamefont {W.}~\bibnamefont
  {Greiner}},\ }\href {\doibase 10.1103/PhysRevC.85.034615} {\bibfield
  {journal} {\bibinfo  {journal} {Phys. Rev. C}\ }\textbf {\bibinfo {volume}
  {85}},\ \bibinfo {pages} {034615} (\bibinfo {year} {2012})}\BibitemShut
  {NoStop}%
\bibitem [{\citenamefont {Poenaru}\ \emph {et~al.}(2013)\citenamefont
  {Poenaru}, \citenamefont {Gherghescu},\ and\ \citenamefont
  {Greiner}}]{Poenaru2013}%
  \BibitemOpen
  \bibfield  {author} {\bibinfo {author} {\bibfnamefont {D.~N.}\ \bibnamefont
  {Poenaru}}, \bibinfo {author} {\bibfnamefont {R.~A.}\ \bibnamefont
  {Gherghescu}}, \ and\ \bibinfo {author} {\bibfnamefont {W.}~\bibnamefont
  {Greiner}},\ }\href {http://stacks.iop.org/1742-6596/436/i=1/a=012056}
  {\bibfield  {journal} {\bibinfo  {journal} {J. Phys. Conf. Ser.}\ }\textbf
  {\bibinfo {volume} {436}},\ \bibinfo {pages} {012056} (\bibinfo {year}
  {2013})}\BibitemShut {NoStop}%
\bibitem [{\citenamefont {Poenaru}\ \emph {et~al.}(2015)\citenamefont
  {Poenaru}, \citenamefont {Gherghescu}, \citenamefont {Greiner},\ and\
  \citenamefont {Shakib}}]{Poenaru2015}%
  \BibitemOpen
  \bibfield  {author} {\bibinfo {author} {\bibfnamefont {D.~N.}\ \bibnamefont
  {Poenaru}}, \bibinfo {author} {\bibfnamefont {R.~A.}\ \bibnamefont
  {Gherghescu}}, \bibinfo {author} {\bibfnamefont {W.}~\bibnamefont {Greiner}},
  \ and\ \bibinfo {author} {\bibfnamefont {N.~S.}\ \bibnamefont {Shakib}},\
  }in\ \href {\doibase 10.1007/978-3-319-10199-6_13} {\emph {\bibinfo
  {booktitle} {Nuclear Physics: Present and Future}}}\ (\bibinfo  {publisher}
  {Springer International Publishing},\ \bibinfo {year} {2015})\ p.\ \bibinfo
  {pages} {131}\BibitemShut {NoStop}%
\bibitem [{\citenamefont {Poenaru}\ and\ \citenamefont
  {Gherghescu}(2018)}]{Poenaru2018}%
  \BibitemOpen
  \bibfield  {author} {\bibinfo {author} {\bibfnamefont {D.~N.}\ \bibnamefont
  {Poenaru}}\ and\ \bibinfo {author} {\bibfnamefont {R.~A.}\ \bibnamefont
  {Gherghescu}},\ }\href {\doibase 10.1103/PhysRevC.97.044621} {\bibfield
  {journal} {\bibinfo  {journal} {Phys. Rev. C}\ }\textbf {\bibinfo {volume}
  {97}},\ \bibinfo {pages} {044621} (\bibinfo {year} {2018})}\BibitemShut
  {NoStop}%
\bibitem [{\citenamefont {Santhosh}\ and\ \citenamefont
  {Nithya}(2018)}]{Santhosh2018}%
  \BibitemOpen
  \bibfield  {author} {\bibinfo {author} {\bibfnamefont {K.~P.}\ \bibnamefont
  {Santhosh}}\ and\ \bibinfo {author} {\bibfnamefont {C.}~\bibnamefont
  {Nithya}},\ }\href {\doibase 10.1103/PhysRevC.97.064616} {\bibfield
  {journal} {\bibinfo  {journal} {Phys. Rev. C}\ }\textbf {\bibinfo {volume}
  {97}},\ \bibinfo {pages} {064616} (\bibinfo {year} {2018})}\BibitemShut
  {NoStop}%
\bibitem [{\citenamefont {Zhang}\ and\ \citenamefont {Wang}(2018)}]{Zhang2018}%
  \BibitemOpen
  \bibfield  {author} {\bibinfo {author} {\bibfnamefont {Y.~L.}\ \bibnamefont
  {Zhang}}\ and\ \bibinfo {author} {\bibfnamefont {Y.~Z.}\ \bibnamefont
  {Wang}},\ }\href {\doibase 10.1103/PhysRevC.97.014318} {\bibfield  {journal}
  {\bibinfo  {journal} {Phys. Rev. C}\ }\textbf {\bibinfo {volume} {97}},\
  \bibinfo {pages} {014318} (\bibinfo {year} {2018})}\BibitemShut {NoStop}%
\bibitem [{\citenamefont {Warda}\ and\ \citenamefont
  {Robledo}(2011)}]{Warda2011}%
  \BibitemOpen
  \bibfield  {author} {\bibinfo {author} {\bibfnamefont {M.}~\bibnamefont
  {Warda}}\ and\ \bibinfo {author} {\bibfnamefont {L.~M.}\ \bibnamefont
  {Robledo}},\ }\href {\doibase 10.1103/PhysRevC.84.044608} {\bibfield
  {journal} {\bibinfo  {journal} {Phys. Rev. C}\ }\textbf {\bibinfo {volume}
  {84}},\ \bibinfo {pages} {044608} (\bibinfo {year} {2011})}\BibitemShut
  {NoStop}%
\bibitem [{\citenamefont {Sadhukhan}\ \emph {et~al.}(2013)\citenamefont
  {Sadhukhan}, \citenamefont {Mazurek}, \citenamefont {Baran}, \citenamefont
  {Dobaczewski}, \citenamefont {Nazarewicz},\ and\ \citenamefont
  {Sheikh}}]{Sadhukhan2013}%
  \BibitemOpen
  \bibfield  {author} {\bibinfo {author} {\bibfnamefont {J.}~\bibnamefont
  {Sadhukhan}}, \bibinfo {author} {\bibfnamefont {K.}~\bibnamefont {Mazurek}},
  \bibinfo {author} {\bibfnamefont {A.}~\bibnamefont {Baran}}, \bibinfo
  {author} {\bibfnamefont {J.}~\bibnamefont {Dobaczewski}}, \bibinfo {author}
  {\bibfnamefont {W.}~\bibnamefont {Nazarewicz}}, \ and\ \bibinfo {author}
  {\bibfnamefont {J.~A.}\ \bibnamefont {Sheikh}},\ }\href {\doibase
  10.1103/PhysRevC.88.064314} {\bibfield  {journal} {\bibinfo  {journal} {Phys.
  Rev. C}\ }\textbf {\bibinfo {volume} {88}},\ \bibinfo {pages} {064314}
  (\bibinfo {year} {2013})}\BibitemShut {NoStop}%
\bibitem [{\citenamefont {Baran}\ \emph {et~al.}(1981)\citenamefont {Baran},
  \citenamefont {Pomorski}, \citenamefont {{\L}ukasiak},\ and\ \citenamefont
  {Sobiczewski}}]{Baran1981}%
  \BibitemOpen
  \bibfield  {author} {\bibinfo {author} {\bibfnamefont {A.}~\bibnamefont
  {Baran}}, \bibinfo {author} {\bibfnamefont {K.}~\bibnamefont {Pomorski}},
  \bibinfo {author} {\bibfnamefont {A.}~\bibnamefont {{\L}ukasiak}}, \ and\
  \bibinfo {author} {\bibfnamefont {A.}~\bibnamefont {Sobiczewski}},\ }\href
  {\doibase https://doi.org/10.1016/0375-9474(81)90471-1} {\bibfield  {journal}
  {\bibinfo  {journal} {Nucl. Phys. A}\ }\textbf {\bibinfo {volume} {361}},\
  \bibinfo {pages} {83 } (\bibinfo {year} {1981})}\BibitemShut {NoStop}%
\bibitem [{\citenamefont {Baran}(1978)}]{Baran1978}%
  \BibitemOpen
  \bibfield  {author} {\bibinfo {author} {\bibfnamefont {A.}~\bibnamefont
  {Baran}},\ }\href {\doibase 10.1016/0370-2693(78)90085-0} {\bibfield
  {journal} {\bibinfo  {journal} {Phys. Lett. B}\ }\textbf {\bibinfo {volume}
  {76}},\ \bibinfo {pages} {8} (\bibinfo {year} {1978})}\BibitemShut {NoStop}%
\bibitem [{\citenamefont {Schunck}\ \emph {et~al.}(2015)\citenamefont
  {Schunck}, \citenamefont {McDonnell}, \citenamefont {Sarich}, \citenamefont
  {Wild},\ and\ \citenamefont {Higdon}}]{Schunck2015}%
  \BibitemOpen
  \bibfield  {author} {\bibinfo {author} {\bibfnamefont {N.}~\bibnamefont
  {Schunck}}, \bibinfo {author} {\bibfnamefont {J.~D.}\ \bibnamefont
  {McDonnell}}, \bibinfo {author} {\bibfnamefont {J.}~\bibnamefont {Sarich}},
  \bibinfo {author} {\bibfnamefont {S.~M.}\ \bibnamefont {Wild}}, \ and\
  \bibinfo {author} {\bibfnamefont {D.}~\bibnamefont {Higdon}},\ }\href
  {http://stacks.iop.org/0954-3899/42/i=3/a=034024} {\bibfield  {journal}
  {\bibinfo  {journal} {J. Phys. G}\ }\textbf {\bibinfo {volume} {42}},\
  \bibinfo {pages} {34024} (\bibinfo {year} {2015})}\BibitemShut {NoStop}%
\bibitem [{\citenamefont {Bartel}\ \emph {et~al.}(1982)\citenamefont {Bartel},
  \citenamefont {Quentin}, \citenamefont {Brack}, \citenamefont {Guet},\ and\
  \citenamefont {H{\aa}kansson}}]{Bartel1982}%
  \BibitemOpen
  \bibfield  {author} {\bibinfo {author} {\bibfnamefont {J.}~\bibnamefont
  {Bartel}}, \bibinfo {author} {\bibfnamefont {P.}~\bibnamefont {Quentin}},
  \bibinfo {author} {\bibfnamefont {M.}~\bibnamefont {Brack}}, \bibinfo
  {author} {\bibfnamefont {C.}~\bibnamefont {Guet}}, \ and\ \bibinfo {author}
  {\bibfnamefont {H.-B.}\ \bibnamefont {H{\aa}kansson}},\ }\href {\doibase
  https://doi.org/10.1016/0375-9474(82)90403-1} {\bibfield  {journal} {\bibinfo
   {journal} {Nucl. Phys. A}\ }\textbf {\bibinfo {volume} {386}},\ \bibinfo
  {pages} {79 } (\bibinfo {year} {1982})}\BibitemShut {NoStop}%
\bibitem [{\citenamefont {Berger}\ \emph {et~al.}(1989)\citenamefont {Berger},
  \citenamefont {Girod},\ and\ \citenamefont {Gogny}}]{Berger1989}%
  \BibitemOpen
  \bibfield  {author} {\bibinfo {author} {\bibfnamefont {J.}~\bibnamefont
  {Berger}}, \bibinfo {author} {\bibfnamefont {M.}~\bibnamefont {Girod}}, \
  and\ \bibinfo {author} {\bibfnamefont {D.}~\bibnamefont {Gogny}},\ }\href
  {\doibase https://doi.org/10.1016/0375-9474(89)90656-8} {\bibfield  {journal}
  {\bibinfo  {journal} {Nucl. Phys. A}\ }\textbf {\bibinfo {volume} {502}},\
  \bibinfo {pages} {85 } (\bibinfo {year} {1989})}\BibitemShut {NoStop}%
\bibitem [{\citenamefont {Sadhukhan}\ \emph {et~al.}(2014)\citenamefont
  {Sadhukhan}, \citenamefont {Dobaczewski}, \citenamefont {Nazarewicz},
  \citenamefont {Sheikh},\ and\ \citenamefont {Baran}}]{Sadhukhan2014}%
  \BibitemOpen
  \bibfield  {author} {\bibinfo {author} {\bibfnamefont {J.}~\bibnamefont
  {Sadhukhan}}, \bibinfo {author} {\bibfnamefont {J.}~\bibnamefont
  {Dobaczewski}}, \bibinfo {author} {\bibfnamefont {W.}~\bibnamefont
  {Nazarewicz}}, \bibinfo {author} {\bibfnamefont {J.~A.}\ \bibnamefont
  {Sheikh}}, \ and\ \bibinfo {author} {\bibfnamefont {A.}~\bibnamefont
  {Baran}},\ }\href {\doibase 10.1103/PhysRevC.90.061304} {\bibfield  {journal}
  {\bibinfo  {journal} {Phys. Rev. C}\ }\textbf {\bibinfo {volume} {90}},\
  \bibinfo {pages} {061304} (\bibinfo {year} {2014})}\BibitemShut {NoStop}%
\bibitem [{\citenamefont {Zhao}\ \emph {et~al.}(2016)\citenamefont {Zhao},
  \citenamefont {Lu}, \citenamefont {Nik\ifmmode \check{s}\else
  \v{s}\fi{}i\ifmmode~\acute{c}\else \'{c}\fi{}}, \citenamefont {Vretenar},\
  and\ \citenamefont {Zhou}}]{Zhao2016}%
  \BibitemOpen
  \bibfield  {author} {\bibinfo {author} {\bibfnamefont {J.}~\bibnamefont
  {Zhao}}, \bibinfo {author} {\bibfnamefont {B.-N.}\ \bibnamefont {Lu}},
  \bibinfo {author} {\bibfnamefont {T.}~\bibnamefont {Nik\ifmmode
  \check{s}\else \v{s}\fi{}i\ifmmode~\acute{c}\else \'{c}\fi{}}}, \bibinfo
  {author} {\bibfnamefont {D.}~\bibnamefont {Vretenar}}, \ and\ \bibinfo
  {author} {\bibfnamefont {S.-G.}\ \bibnamefont {Zhou}},\ }\href {\doibase
  10.1103/PhysRevC.93.044315} {\bibfield  {journal} {\bibinfo  {journal} {Phys.
  Rev. C}\ }\textbf {\bibinfo {volume} {93}},\ \bibinfo {pages} {044315}
  (\bibinfo {year} {2016})}\BibitemShut {NoStop}%
\bibitem [{\citenamefont {Schunck}\ \emph {et~al.}(2017)\citenamefont {Schunck}
  \emph {et~al.}}]{Schunck2017}%
  \BibitemOpen
  \bibfield  {author} {\bibinfo {author} {\bibfnamefont {N.}~\bibnamefont
  {Schunck}} \emph {et~al.},\ }\href {\doibase
  https://doi.org/10.1016/j.cpc.2017.03.007} {\bibfield  {journal} {\bibinfo
  {journal} {Comput. Phys. Commun.}\ }\textbf {\bibinfo {volume} {216}},\
  \bibinfo {pages} {145 } (\bibinfo {year} {2017})}\BibitemShut {NoStop}%
\bibitem [{\citenamefont {Mcdonnell}\ \emph {et~al.}(2014)\citenamefont
  {Mcdonnell}, \citenamefont {Nazarewicz}, \citenamefont {Sheikh},
  \citenamefont {Staszczak},\ and\ \citenamefont {Warda}}]{Mcdonnell2014}%
  \BibitemOpen
  \bibfield  {author} {\bibinfo {author} {\bibfnamefont {J.~D.}\ \bibnamefont
  {Mcdonnell}}, \bibinfo {author} {\bibfnamefont {W.}~\bibnamefont
  {Nazarewicz}}, \bibinfo {author} {\bibfnamefont {J.~A.}\ \bibnamefont
  {Sheikh}}, \bibinfo {author} {\bibfnamefont {A.}~\bibnamefont {Staszczak}}, \
  and\ \bibinfo {author} {\bibfnamefont {M.}~\bibnamefont {Warda}},\ }\href
  {\doibase 10.1103/PhysRevC.90.021302} {\bibfield  {journal} {\bibinfo
  {journal} {Phys. Rev. C}\ }\textbf {\bibinfo {volume} {021302}},\ \bibinfo
  {pages} {18} (\bibinfo {year} {2014})}\BibitemShut {NoStop}%
\bibitem [{Rob()}]{Robledo2002}%
  \BibitemOpen
  \href@noop {} {}\bibinfo {note} {Luis M. Robledo, code HFBaxial, 2002,
  \url{http://gamma.ft.uam.es/robledo/Downloads.html}}\BibitemShut {NoStop}%
\bibitem [{\citenamefont {Baran}\ \emph {et~al.}(2011)\citenamefont {Baran},
  \citenamefont {Sheikh}, \citenamefont {Dobaczewski}, \citenamefont
  {Nazarewicz},\ and\ \citenamefont {Staszczak}}]{Baran2011}%
  \BibitemOpen
  \bibfield  {author} {\bibinfo {author} {\bibfnamefont {A.}~\bibnamefont
  {Baran}}, \bibinfo {author} {\bibfnamefont {J.~A.}\ \bibnamefont {Sheikh}},
  \bibinfo {author} {\bibfnamefont {J.}~\bibnamefont {Dobaczewski}}, \bibinfo
  {author} {\bibfnamefont {W.}~\bibnamefont {Nazarewicz}}, \ and\ \bibinfo
  {author} {\bibfnamefont {A.}~\bibnamefont {Staszczak}},\ }\href {\doibase
  10.1103/PhysRevC.84.054321} {\bibfield  {journal} {\bibinfo  {journal} {Phys.
  Rev. C}\ }\textbf {\bibinfo {volume} {84}},\ \bibinfo {pages} {054321}
  (\bibinfo {year} {2011})}\BibitemShut {NoStop}%
\bibitem [{\citenamefont {Giuliani}\ and\ \citenamefont
  {Robledo}(2018)}]{Giuliani2018b}%
  \BibitemOpen
  \bibfield  {author} {\bibinfo {author} {\bibfnamefont {S.~A.}\ \bibnamefont
  {Giuliani}}\ and\ \bibinfo {author} {\bibfnamefont {L.~M.}\ \bibnamefont
  {Robledo}},\ }\href {\doibase 10.1016/j.physletb.2018.10.045} {\bibfield
  {journal} {\bibinfo  {journal} {Phys. Lett. B}\ }\textbf {\bibinfo {volume}
  {787}},\ \bibinfo {pages} {134} (\bibinfo {year} {2018})}\BibitemShut
  {NoStop}%
\bibitem [{\citenamefont {Usang}\ \emph {et~al.}(2017)\citenamefont {Usang},
  \citenamefont {Ivanyuk}, \citenamefont {Ishizuka},\ and\ \citenamefont
  {Chiba}}]{usang2017}%
  \BibitemOpen
  \bibfield  {author} {\bibinfo {author} {\bibfnamefont {M.~D.}\ \bibnamefont
  {Usang}}, \bibinfo {author} {\bibfnamefont {F.~A.}\ \bibnamefont {Ivanyuk}},
  \bibinfo {author} {\bibfnamefont {C.}~\bibnamefont {Ishizuka}}, \ and\
  \bibinfo {author} {\bibfnamefont {S.}~\bibnamefont {Chiba}},\ }\href
  {\doibase 10.1103/PhysRevC.96.064617} {\bibfield  {journal} {\bibinfo
  {journal} {Phys. Rev. C}\ }\textbf {\bibinfo {volume} {96}},\ \bibinfo
  {pages} {064617} (\bibinfo {year} {2017})}\BibitemShut {NoStop}%
\bibitem [{\citenamefont {Ishizuka}\ \emph {et~al.}(2017)\citenamefont
  {Ishizuka}, \citenamefont {Usang}, \citenamefont {Ivanyuk}, \citenamefont
  {Maruhn}, \citenamefont {Nishio},\ and\ \citenamefont
  {Chiba}}]{ishizuka2017}%
  \BibitemOpen
  \bibfield  {author} {\bibinfo {author} {\bibfnamefont {C.}~\bibnamefont
  {Ishizuka}}, \bibinfo {author} {\bibfnamefont {M.~D.}\ \bibnamefont {Usang}},
  \bibinfo {author} {\bibfnamefont {F.~A.}\ \bibnamefont {Ivanyuk}}, \bibinfo
  {author} {\bibfnamefont {J.~A.}\ \bibnamefont {Maruhn}}, \bibinfo {author}
  {\bibfnamefont {K.}~\bibnamefont {Nishio}}, \ and\ \bibinfo {author}
  {\bibfnamefont {S.}~\bibnamefont {Chiba}},\ }\href {\doibase
  10.1103/PhysRevC.96.064616} {\bibfield  {journal} {\bibinfo  {journal} {Phys.
  Rev. C}\ }\textbf {\bibinfo {volume} {96}},\ \bibinfo {pages} {064616}
  (\bibinfo {year} {2017})}\BibitemShut {NoStop}%
\bibitem [{\citenamefont {Heenen}\ \emph {et~al.}(2015)\citenamefont {Heenen},
  \citenamefont {Skalski}, \citenamefont {Staszczak},\ and\ \citenamefont
  {Vretenar}}]{Heenen2015}%
  \BibitemOpen
  \bibfield  {author} {\bibinfo {author} {\bibfnamefont {P.-H.}\ \bibnamefont
  {Heenen}}, \bibinfo {author} {\bibfnamefont {J.}~\bibnamefont {Skalski}},
  \bibinfo {author} {\bibfnamefont {A.}~\bibnamefont {Staszczak}}, \ and\
  \bibinfo {author} {\bibfnamefont {D.}~\bibnamefont {Vretenar}},\ }\href
  {\doibase 10.1016/j.nuclphysa.2015.07.016} {\bibfield  {journal} {\bibinfo
  {journal} {Nucl. Phys. A}\ }\textbf {\bibinfo {volume} {944}},\ \bibinfo
  {pages} {415} (\bibinfo {year} {2015})}\BibitemShut {NoStop}%
\bibitem [{NUD()}]{NUDAT}%
  \BibitemOpen
  \href@noop {} {}\bibinfo {note} {Interactive Chart of Nuclides, NuDat 2.7
  \url{https://www.nndc.bnl.gov/nudat2/}}\BibitemShut {NoStop}%
\bibitem [{\citenamefont {{Bulgac}}\ \emph {et~al.}(2018)\citenamefont
  {{Bulgac}}, \citenamefont {{Jin}}, \citenamefont {{Roche}}, \citenamefont
  {{Schunck}},\ and\ \citenamefont {{Stetcu}}}]{Bulgac2018}%
  \BibitemOpen
  \bibfield  {author} {\bibinfo {author} {\bibfnamefont {A.}~\bibnamefont
  {{Bulgac}}}, \bibinfo {author} {\bibfnamefont {S.}~\bibnamefont {{Jin}}},
  \bibinfo {author} {\bibfnamefont {K.}~\bibnamefont {{Roche}}}, \bibinfo
  {author} {\bibfnamefont {N.}~\bibnamefont {{Schunck}}}, \ and\ \bibinfo
  {author} {\bibfnamefont {I.}~\bibnamefont {{Stetcu}}},\ }\href@noop {}
  {\bibfield  {journal} {\bibinfo  {journal} {ArXiv e-prints}\ } (\bibinfo
  {year} {2018})},\ \Eprint {http://arxiv.org/abs/1806.00694} {arXiv:1806.00694
  [nucl-th]} \BibitemShut {NoStop}%
\bibitem [{\citenamefont {Randrup}\ \emph {et~al.}(2011)\citenamefont
  {Randrup}, \citenamefont {M\"oller},\ and\ \citenamefont
  {Sierk}}]{Randrup2011}%
  \BibitemOpen
  \bibfield  {author} {\bibinfo {author} {\bibfnamefont {J.}~\bibnamefont
  {Randrup}}, \bibinfo {author} {\bibfnamefont {P.}~\bibnamefont {M\"oller}}, \
  and\ \bibinfo {author} {\bibfnamefont {A.~J.}\ \bibnamefont {Sierk}},\ }\href
  {\doibase 10.1103/PhysRevC.84.034613} {\bibfield  {journal} {\bibinfo
  {journal} {Phys. Rev. C}\ }\textbf {\bibinfo {volume} {84}},\ \bibinfo
  {pages} {034613} (\bibinfo {year} {2011})}\BibitemShut {NoStop}%
\bibitem [{\citenamefont {Sierk}(2017)}]{Sierk2017}%
  \BibitemOpen
  \bibfield  {author} {\bibinfo {author} {\bibfnamefont {A.~J.}\ \bibnamefont
  {Sierk}},\ }\href {\doibase 10.1103/PhysRevC.96.034603} {\bibfield  {journal}
  {\bibinfo  {journal} {Phys. Rev. C}\ }\textbf {\bibinfo {volume} {96}},\
  \bibinfo {pages} {034603} (\bibinfo {year} {2017})}\BibitemShut {NoStop}%
\bibitem [{\citenamefont {Sadhukhan}\ \emph {et~al.}(2017)\citenamefont
  {Sadhukhan}, \citenamefont {Zhang}, \citenamefont {Nazarewicz},\ and\
  \citenamefont {Schunck}}]{Sadhukhan2017}%
  \BibitemOpen
  \bibfield  {author} {\bibinfo {author} {\bibfnamefont {J.}~\bibnamefont
  {Sadhukhan}}, \bibinfo {author} {\bibfnamefont {C.}~\bibnamefont {Zhang}},
  \bibinfo {author} {\bibfnamefont {W.}~\bibnamefont {Nazarewicz}}, \ and\
  \bibinfo {author} {\bibfnamefont {N.}~\bibnamefont {Schunck}},\ }\href
  {\doibase 10.1103/PhysRevC.96.061301} {\bibfield  {journal} {\bibinfo
  {journal} {Phys. Rev. C}\ }\textbf {\bibinfo {volume} {061301}},\ \bibinfo
  {pages} {1} (\bibinfo {year} {2017})}\BibitemShut {NoStop}%
\bibitem [{\citenamefont {Aritomo}\ \emph {et~al.}(2014)\citenamefont
  {Aritomo}, \citenamefont {Chiba},\ and\ \citenamefont
  {Ivanyuk}}]{Arimoto2014}%
  \BibitemOpen
  \bibfield  {author} {\bibinfo {author} {\bibfnamefont {Y.}~\bibnamefont
  {Aritomo}}, \bibinfo {author} {\bibfnamefont {S.}~\bibnamefont {Chiba}}, \
  and\ \bibinfo {author} {\bibfnamefont {F.}~\bibnamefont {Ivanyuk}},\ }\href
  {\doibase 10.1103/PhysRevC.90.054609} {\bibfield  {journal} {\bibinfo
  {journal} {Phys. Rev. C}\ }\textbf {\bibinfo {volume} {90}},\ \bibinfo
  {pages} {054609} (\bibinfo {year} {2014})}\BibitemShut {NoStop}%
\bibitem [{\citenamefont {Zhang}\ \emph {et~al.}(2016)\citenamefont {Zhang},
  \citenamefont {Schuetrumpf},\ and\ \citenamefont {Nazarewicz}}]{Zhang2016}%
  \BibitemOpen
  \bibfield  {author} {\bibinfo {author} {\bibfnamefont {C.~L.}\ \bibnamefont
  {Zhang}}, \bibinfo {author} {\bibfnamefont {B.}~\bibnamefont {Schuetrumpf}},
  \ and\ \bibinfo {author} {\bibfnamefont {W.}~\bibnamefont {Nazarewicz}},\
  }\href {\doibase 10.1103/PhysRevC.94.064323} {\bibfield  {journal} {\bibinfo
  {journal} {Phys. Rev. C}\ }\textbf {\bibinfo {volume} {94}},\ \bibinfo
  {pages} {1} (\bibinfo {year} {2016})}\BibitemShut {NoStop}%
\end{thebibliography}%

\end{document}